\documentclass[twocolumn,10pt]{article}

\usepackage[a4paper,margin=0.7in]{geometry}
\usepackage[english]{babel}
\usepackage{float}
\usepackage[dvipsnames]{xcolor}
\usepackage{graphicx}
\usepackage{amsmath, amsthm, amssymb}
\usepackage{amsfonts}
\usepackage{romannum}
\usepackage{tikz}
\usepackage{titlesec}
\usepackage{abstract}
\usepackage[numbers,sort&compress]{natbib}
\usepackage{caption}
\usepackage{subcaption}
\usepackage{mathrsfs} 
\usepackage{upgreek}
\usepackage[T1]{fontenc}
\usepackage{lmodern}
\usepackage{booktabs}
\usepackage{tabularx}
\usepackage{siunitx}

\usepackage[
    colorlinks=true,
    linkcolor=red,
    filecolor=blue,
    urlcolor=blue,
    citecolor=blue,
    breaklinks=true
]{hyperref}

\newcommand{\orcid}[1]{%
  \,\href{https://orcid.org/#1}{\raisebox{-0.4pt}{\includegraphics[width=8pt]{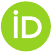}}}%
}

\renewcommand{\thesection}{\Roman{section}}
\renewcommand{\thesubsection}{\Roman{section}.\Alph{subsection}}
\newcommand{\ts}[1]{\textsuperscript{#1}}

\titleformat{\section}
  {\large\bfseries\fontsize{10}{12}\selectfont\MakeUppercase}
  {\thesection.}{0.5em}{}

\titleformat{\subsection}
  {\normalsize\bfseries}
  {\thesubsection.}{0.5em}{}

\title{%
\vspace{-1.5em}
\textbf{\fontsize{16pt}{16pt}\selectfont\bfseries Quantum metrology with undetected mid-infrared photons for applied non-destructive testing}
\vspace{-0.5em}
}

\author{%
\parbox{0.96\textwidth}{\centering
\normalsize
Paul Gattinger\orcid{0000-0002-0120-9100}$^{1}$\quad
Andreas W. Schell\orcid{0000-0003-0849-9558}$^{2}$\\[0.25em]
Maja Buchegger$^{1}$\quad
Markus Brandstetter\orcid{0000-0002-8679-8097}$^{1}$\quad
Ivan Zorin\orcid{0000-0002-2089-5716}$^{1,*}$\\[0.8em]
\footnotesize\itshape
$^{1}$Research Center for Non-Destructive Testing, Science Park 2, Altenberger Str. 69, 4040 Linz, Austria\\
$^{2}$Division of Light-Matter-Interaction, Johannes Kepler University, Altenberger Str. 69, 4040 Linz, Austria\\[0.6em]
\footnotesize\normalfont
$^{*}$Corresponding authors: \href{mailto:paul.gattinger@recendt.at}{paul.gattinger@recendt.at} and \href{mailto:ivan.zorin@recendt.at}{ivan.zorin@recendt.at}\\[0.6em]
https://doi.org/10.1007/s00502-026-01428-3\\ \vspace{5pt}
\footnotesize
\today
}
}
\date{}

\begin{document}
\twocolumn[
\maketitle
\vspace{-2em}

\begin{onecolabstract} 
Metrology with undetected photons is an emerging technique that leverages quantum effects and photon correlations (entanglement) to retrieve valuable information in a target spectral range (e.g., mid-infrared, mid-IR) using measurements in an easily accessible domain (e.g., visible, near-IR). The underlying quantum process of spontaneous parametric down-conversion (SPDC) is utilized to generate non-degenerate correlated signal and idler photons to serve as detection and probing photons, respectively. Sensing with undetected photons enables important advantages, such as ultra-low probe powers, room-temperature operation, and shot-noise-limited detection. In this contribution, we apply a quantum nonlinear interferometer based on an SPDC source to perform applied mid-IR spectroscopy, mid-IR microscopy, and mid-IR optical coherence tomography (OCT) as among the most promising techniques for quantum-based routine non-destructive testing. Moreover, we characterize the system, benchmark it against classical systems, and provide a prospective outlook for this new technology. 
\end{onecolabstract}
\vspace{0.5em}
]\saythanks

% \section{Introduction and approach of the problem}\label{sec1}
\section{Introduction}\label{sec1}

% REMINDER: The subject should be treated in an application-oriented way, to be comprehended not only by few experts, but by a broad public.

% Set-up of the paper: Original papers are to be divided into introduction and approach of the problem, methods and discussion as well as summary and conclusion. 
Metrology with undetected photons has been first introduced in 2014~\cite{Lemos2014} and has since become a topic of extensive research, with many demonstrated applications, ranging from imaging~\cite{Kviatkovsky2020,Pearce2023,GilaberteBasset2023,Placke2026}, infrared (IR) and terahertz (THz) spectroscopy~\cite{Kalashnikov2016,Lindner2022,Kaufmann2022,Lindner2023,Tashima2024,Kaur2024,Gattinger2025,Kutas2020}, to optical coherence tomography (OCT)~\cite{Paterova2018,Vanselow2020}. The principle of this technology is based on nonlinear interferometry and utilizes spontaneous parametric down-conversion (SPDC), where photons of a shorter (pump) wavelength are converted to correlated photon pairs of longer wavelengths. These generated bi-photons can be of vastly different wavelengths (non-degenerate) by design~\cite{Vanselow2019}. The key idea is to use the photons of longer wavelengths (idler, mid-IR) to probe a sample, and the photons with shorter wavelengths (signal, usually in the visible and near-IR) for detection. Due to the correlations within the bi-photons, information about the photons of longer wavelengths (sample-induced absorption, scattering, phase shift) can be read out interferometrically via the near-IR photons. Thus, instead of technically demanding, complex, and expensive mid-IR laser sources and mid-IR detectors, well-developed visible lasers and shot-noise-limited Si-based detectors can be used.

Sensing with undetected photons is an intrinsically interferometric, albeit non-classical, quantum optical technique. Well-established classical non-destructive testing methods can be implemented on this basis and can benefit from this measurement paradigm.

Mid-IR spectroscopy is a well-established method for qualitative and quantitative chemical analysis and non-destructive metrology of organic compounds that display ro-vibrational transitions~\cite{Thompson2018}. Molecules exhibit fundamental absorptions in the mid-IR spectral domain determined by their structure; these absorption bands are unique and can be accessed via Mid-IR spectroscopy. The gold-standard instrumentation in applied Mid-IR spectroscopy is Fourier transform infrared spectroscopy (FTIR). The FTIR method is based on classical low-coherence interferometry; its transition to nonlinear interferometry with non-degenerate SPDC light provides distinct advantages, such as ultra-low probing powers, room-temperature operation, and shot-noise limited detection. The non-classical QFTIR~\cite{Gattinger2025} (quantum FTIR) modality can also be advanced to imaging and microscopy.

OCT is an interferometric non-destructive method for sub-surface morphological imaging~\cite{drexler_optical_2015}, well-established in the near-IR domain. The primary application area of OCT is biomedical imaging, with a smaller share in non-destructive material characterization~\cite{stifter_beyond_2007}. A primary constraint in applied materials research and defectoscopy is limited performance, for instance, due to increased light scattering in porous samples. OCT at longer wavelengths, for example, in the mid-IR domain, is a potential solution~\cite{Su:14}. However, its realization using state-of-the-art mid-IR technologies is technically challenging due to the lack of well-developed, affordable, and broadband IR sources and low noise, fast, uncooled IR detectors. Since mid-IR light is less prone to scattering, OCT in this spectral domain is especially attractive for highly porous materials such as ceramics or paintings and opens up a range of formerly inaccessible applications. OCT with undetected photons provides the necessary degrees of freedom to operate in the mid-IR domain utilizing well-established, cost-effective technologies for detection and photon generation in the visible and near-IR domain. At this point, it is important to emphasize that quantum-based OCT with undetected photons is not to be confused with quantum-OCT, which relies on coincident measurements of bi-photons~\cite{Abouraddy2002} and is not considered in this work.

In this contribution, we demonstrate mid-IR microscopic mapping and QFTIR spectroscopy as well as mid-IR OCT with undetected photons. We review the current state of the techniques, benchmark the performance, identify the benefits of sensing with undetected photons, and discuss prospects for further development of these methods.
%Especially in the near-infrared (near-IR) regime, modern sources and detectors are well developed.  

\section{Methods and discussion}

The core of any measurement device for metrology with undetected photons is a nonlinear interferometer, where passive optical beam splitters are replaced with active parametric amplifiers (i.e., nonlinear crystals), whose crystal structures lead to a nonlinear response to the strength of the applied optical field~\cite{Boyd2020}. If a nonlinear crystal is pumped with a pump laser at frequency $\omega_p$, there is a small probability (usually on the order of $10^{-9}$) that a pump photon could be converted into two photons of lower frequencies $\omega_s$ (signal) and $\omega_i$ (idler). This process is also known as SPDC. The possible frequencies are dictated by the quasi phase matching condition~\cite{Vanselow2019} and the energy conservation law,
\begin{equation}
\omega_p = \omega_s + \omega_i.
\end{equation}
Two of the most commonly used crystal systems for applied mid-IR sensing are periodically poled potassium titanyl phosphate (ppKTP) and periodically poled lithium niobate (ppLN). The periodic poling of the crystal domains can be designed such that the resulting down-converted photons are emitted quasi-collinearly and cover large far-separated spectral bands, as visualized in Fig.~\ref{fig:energy_conservation}. If two spatially separated crystals are pumped sequentially, SPDC could occur in any of the two crystals. 
\begin{figure}[hbt]
\begin{figure}[b!]
\centering
\includegraphics[width=0.49\textwidth]{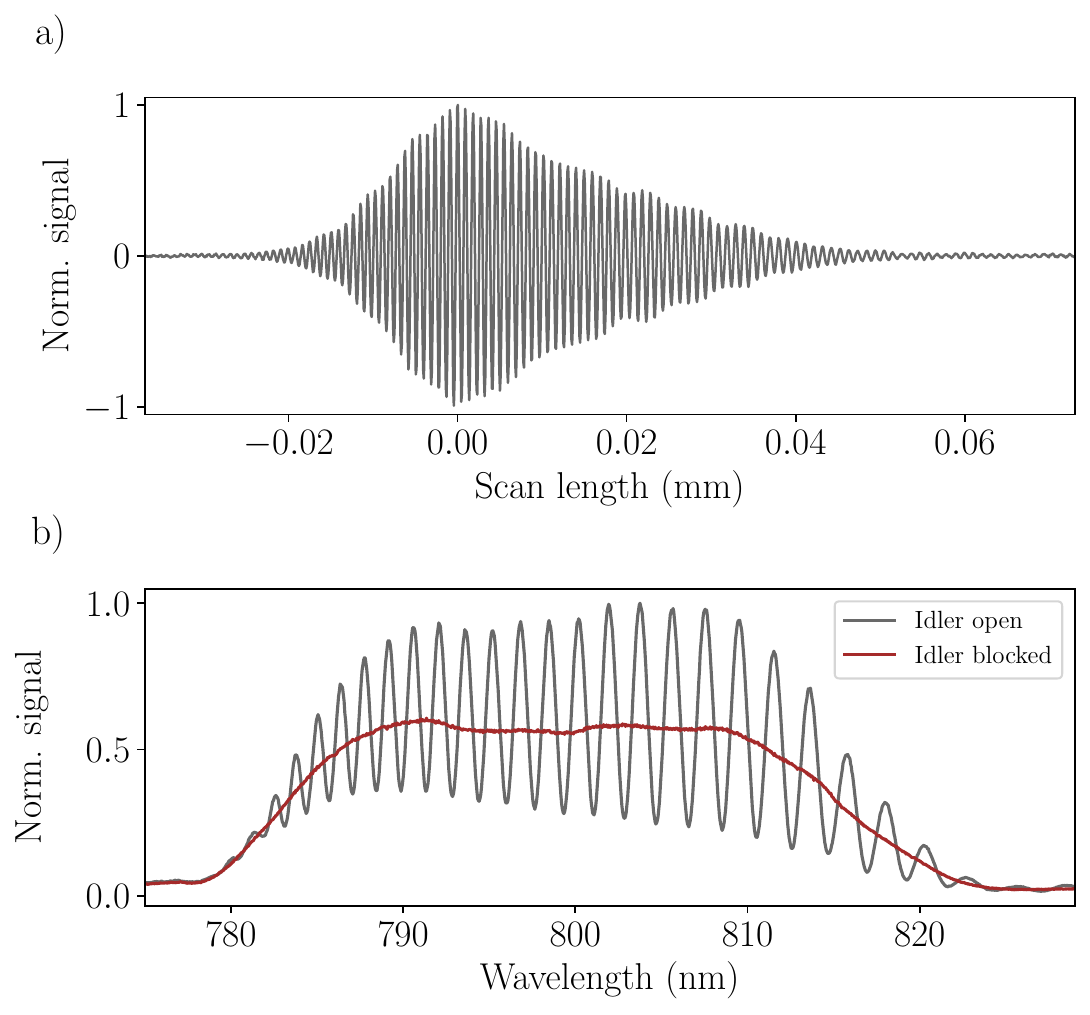}
\caption{Signal photon interferograms corresponding to the measurement modalities. a) high-pass filtered time-domain interferogram of the QFTIR modality (1~s scan time); b) Fourier-domain spectral interferogram (red line) for OCT with undetected photons (100~ms integration time); and signal SPDC profile with idler photons blocked after first SPDC process (gray line).} \label{fig:rawsig}
\end{figure}

\centering
\includegraphics[width=0.45\textwidth,trim={8cm 0cm 7.9cm 9cm},clip]{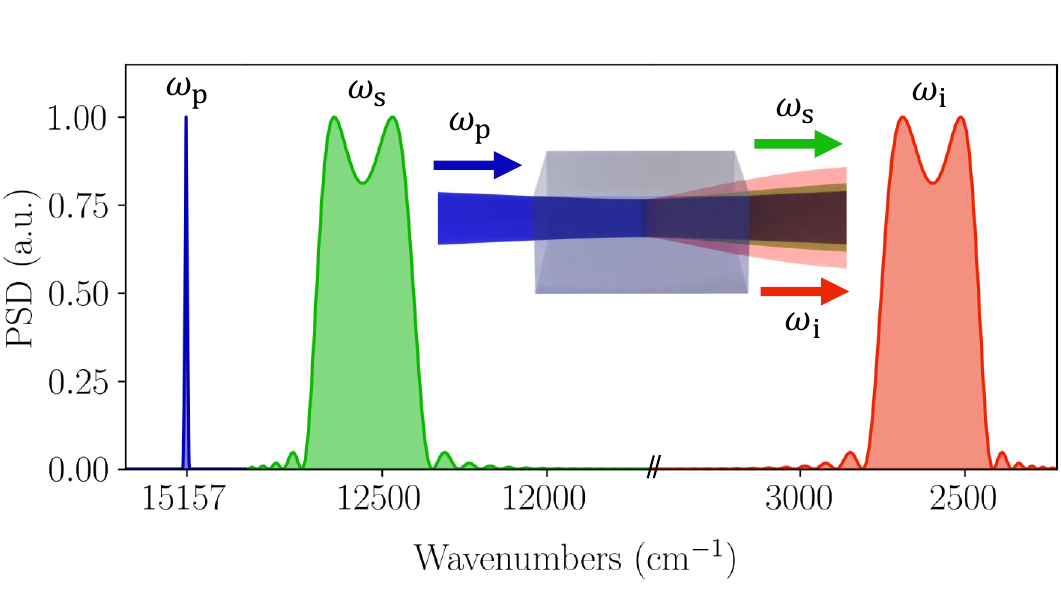}
\caption{Frequency-domain representation: simulated power spectral densities (PSDs) of the narrowband pump $\omega_p$, signal photons $\omega_s$ used for detection and idler photons $\omega_i$ used for probing, all individually normalized to 1 for illustration purposes. The inset depicts SPDC generation in a quasi-phase-matched configuration, where a pump beam propagating through a nonlinear crystal generates signal and idler beams.} \label{fig:energy_conservation}
\end{figure}
If the resulting signal and idler photon paths are overlapped correctly in the second crystal, they become indistinguishable. This is the fundamental condition to observe interference in the signal (and, in principle, in the idler) domain. This phenomenon is also known as induced coherence without induced emission~\cite{Mandel1991,Zou1991}. If, for instance, the cumulative phase of the pump, signal and idler photons, $\Delta \phi = \phi_\mathbf{p} - \phi_\mathbf{s} - \phi_\mathbf{i}$, is changed between the two crystals, then we can observe a change of interference in the signal photon rate $R_s$ according to 
\begin{equation}
R_s = \frac{1+|T|\cos(\Delta \phi + \gamma)}{2},
\end{equation} \label{eqs:interference}
where $|T|$ is the idler transmission through the sample and $\gamma$ is the phase delay introduced by a sample. Eq.~\ref{eqs:interference} shows that, if idler photons are absorbed (or scattered) between the two crystals, the interference observed in the signal photons entirely disappears. Hence, absorbers, scatterers, and phase shifters in the mid-IR (idler domain) can be detected by measuring interference of the signal photons. Figure~\ref{fig:rawsig} shows signal photon interference in the time-domain (for QFTIR measurements, 1~s scan time, scan length cropped for illustrative purposes) and the Fourier-domain (for quantum-based OCT), respectively. A high-pass filter was applied to the time-domain signal in order to remove low-frequency artifacts introduced by the scanning stage. The asymmetry of the interferogram in the empty interferometer comes from the strong unbalanced group velocity dispersion, which is inherent to this type of nonlinear interferometer~\cite{Zorin2026}. 
The Fourier-domain signals for both open and blocked idler arm were acquired with 100~ms integration time. It should be noted, that by blocking the idler photons, the integral over the Fourier-domain spectrum is conserved, since the number of the detected signal photons is constant. 
\begin{figure*}[ht!]
\centering
\begin{tikzpicture}
  \node[anchor=south west,inner sep=0] (image) at (0,0,0) {\includegraphics[width=0.9\textwidth]{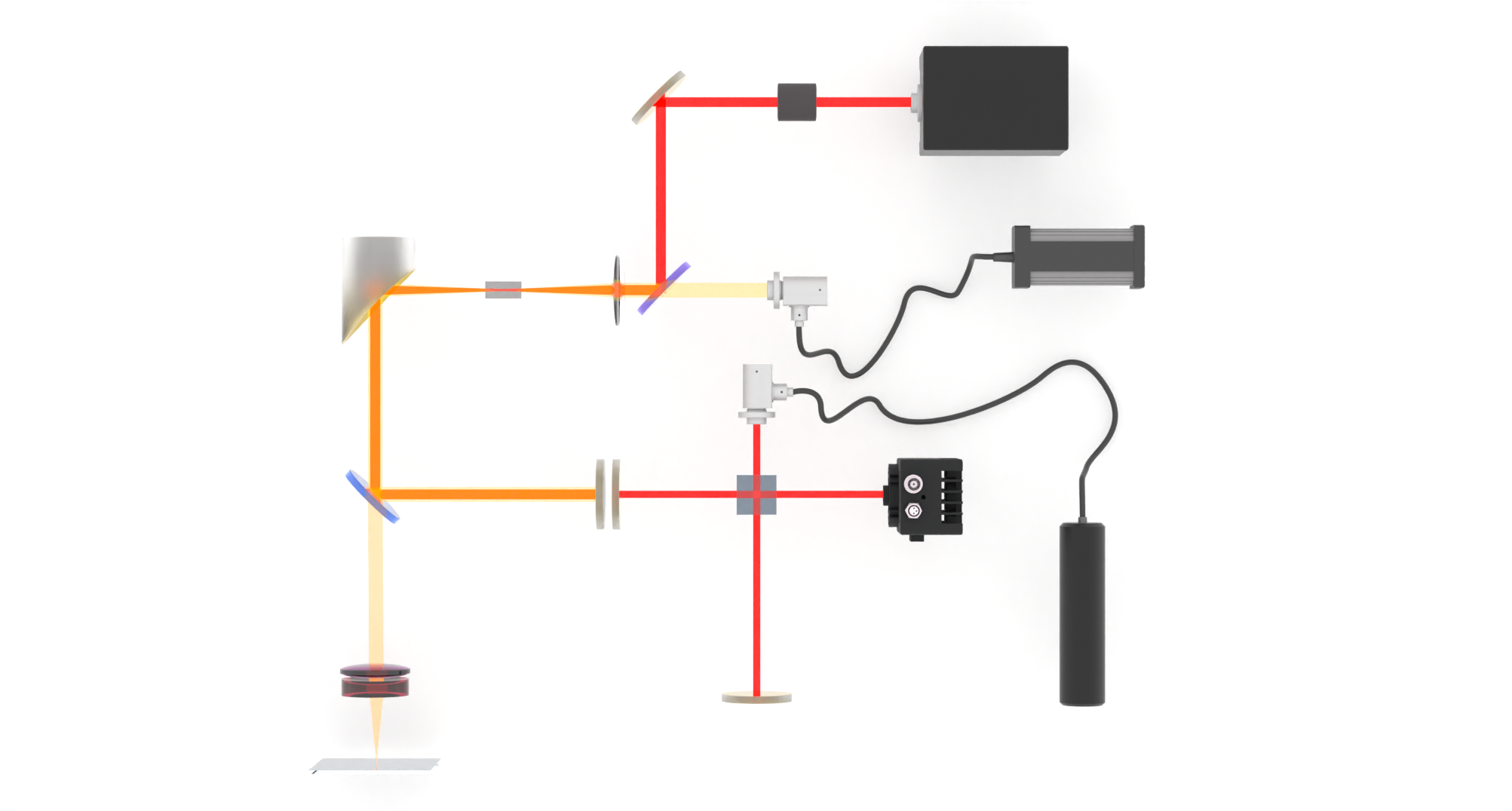}};
  \begin{scope}[x={(image.south east)},y={(image.north west)}]
    %% next four lines will help you to locate the point needed by forming a grid. comment these four lines in the final picture.↓
    % \draw[help lines,xstep=.1,ystep=.1] (0,0) grid (1,1);
    % \draw[help lines,xstep=.05,ystep=.05] (0,0) grid (1,1);
    % \foreach \x in {0,1,...,9} { \node [anchor=north] at (\x/10,0) {0.\x}; }
    % \foreach \y in {0,1,...,9} { \node [anchor=east] at (0,\y/10) {0.\y};}
    %% upto here↑
    \draw (0.655,0.97) node[]{\color{black}\footnotesize Pump laser};
    \draw (0.715,0.745) node[]{\color{black}\footnotesize Detector};
    \draw (0.53,0.92) node[]{\color{black}\footnotesize Isolator};
    \draw (0.423,0.92) node[]{\color{black}\footnotesize AM};
    \draw (0.33,0.68) node[]{\color{black}\footnotesize ppKTP};
    \draw (0.25,0.734) node[]{\color{black}\footnotesize OAPM};
    \draw (0.407,0.565) node[]{\color{black}\footnotesize AC lens};
    \draw (0.47,0.7) node[]{\color{black}\footnotesize CM};
    \draw (0.22,0.37) node[]{\color{black}\footnotesize DM};
    \draw (0.405,0.46) node[]{\color{black}\footnotesize SM};
    %\draw (0.29,0.055) node[]{\color{black}\footnotesize IM};
    \draw (0.475,0.34) node[]{\color{black}\footnotesize BS};
    \draw (0.545,0.14) node[]{\color{black}\footnotesize FM};
    \draw (0.185,0.25) node[]{\color{black}\footnotesize Idler arm};
    \draw (0.32,0.425) node[]{\color{black}\footnotesize Signal arm};
    \draw (0.33,0.16) node[]{\color{black}\footnotesize AC doublet};
    \draw (0.33,0.06) node[]{\color{black}\footnotesize Sample};
    \draw (0.61,0.3) node[]{\color{black}\footnotesize Ref. Detector};
    \draw (0.745,0.24) node[rotate=90]{\color{black}\footnotesize HeNe laser};
    \draw [latex-latex,thick, color={black}] (0.40-0.03, 0.33) -- (0.40+0.03, 0.33);
  \end{scope}
\end{tikzpicture}
\caption{Experimental multimodal system for metrology with undetected photons: nonlinear Michelson-type interferometer; 660~nm CW laser pumps a ppKTP crystal ($\Lambda=20.45$~\textmu m, $L=2.55$~mm), launched via an alignment mirror (AM) and a cold mirror (CM), focused by an achromatic (AC) lens ($f=75$~mm).
The targeted idler spectral band (used for probing) spans 3000~cm\ts{-1} – 2360~cm\ts{-1} (3333~nm – 4237~nm), which corresponds to the signal range from approximately 780~nm to 820~nm (used for detection). The generated signal and idler beams are collimated via an off-axis parabolic mirror (OAPM, $f=75$~mm) and separated by a dichroic mirror (DM). The combined signal and pump arm length is varied by a scanning mirror (SM). For the IR microscopic mapping modality and OCT imaging a $f=50$~mm AC-doublet was used. The reference interferometer is a Michelson interferometer composed of a HeNe laser, a beam splitter (BS) and a fixed mirror (FM).} \label{fig:setup}
\end{figure*}
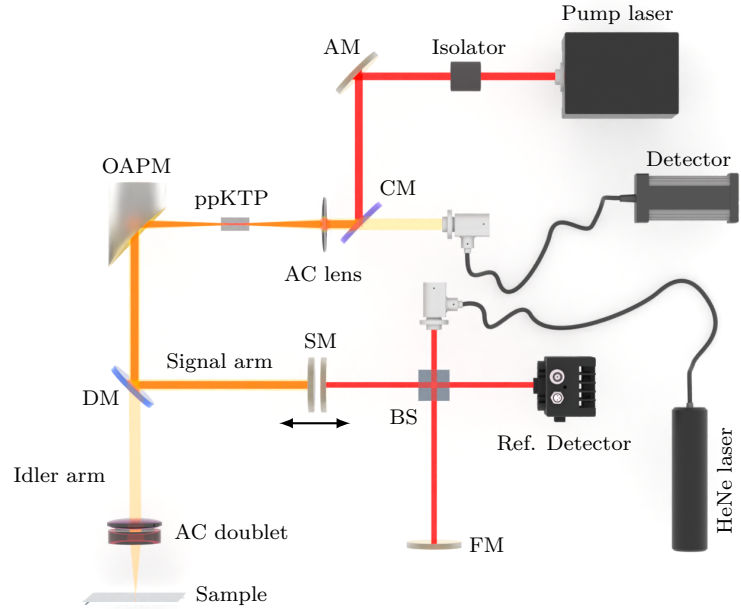
This highlights a major difference to classical (Michelson) interferometry, where blocking the reference arm reduces the signal magnitude by the corresponding beam splitter ratio. 
Despite the above-mentioned peculiarities, the raw signals in QFTIR and quantum-based OCT resemble those of analogous classical systems. This enables the adoption of existing, well-established post-processing algorithms with only minor adjustments.

The generalized schematic of a nonlinear interferometer can be collapsed to a folded geometry, so that one nonlinear crystal is pumped sequentially. Thus, consecutive propagations through two crystals are replaced by forward and backward propagation through the very same crystal. Such a practical experimental setup is shown in Fig.~\ref{fig:setup}. The arrangement is referred to as a nonlinear Michelson-type interferometer~\cite{Chekhova2016}.

The pump laser (Cobolt Flamenco, 660~nm, 500~mW) is reflected by a cold mirror (CM) and then focused by an achromatic (AC) lens ($f=75$~mm) into a ppKTP crystal ($\Lambda=20.45$~\textmu m, $L=2.55$~mm) with a resulting focusing parameter (ratio of crystal length and confocal parameter) of 1.42. An off-axis parabolic mirror (OAPM) is used to collimate the pump and the emerging signal and idler photon beams. A dichroic mirror (DM) reflects the pump and the signal, forming the combined pump and signal arm, and transmits the idler into the idler arm, where a Si-Ge AC doublet with $f=50$~mm is used to focus the idler photons onto the sample. The photons from the combined pump and signal arm are reflected off a scanning mirror (SM); the idler photons that interacted with the sample propagate back to the crystal. Pump, signal, and idler photons are then combined by the DM and pass the ppKTP crystal a second time (paths are overlapped), where pump photons could again be converted to signal and idler photons. The signal photons are coupled out through the CM and detected by either a spectrometer (customized Ocean Optics HR6) or a point detector (Femto OE-200-SI-FC), depending on the measurement modality~\textemdash optical coherence tomography (OCT) with undetected photons or quantum Fourier transform infrared spectroscopy (QFTIR), respectively. The pump photons are reflected towards the optical isolator, where they are absorbed. The idler photons are disregarded, as they are not important for the measurement after the second pass through the crystal. 

\begin{figure*}[htb]
\centering
\includegraphics[width=1\textwidth,trim={1cm 0cm 1cm 0cm},clip]{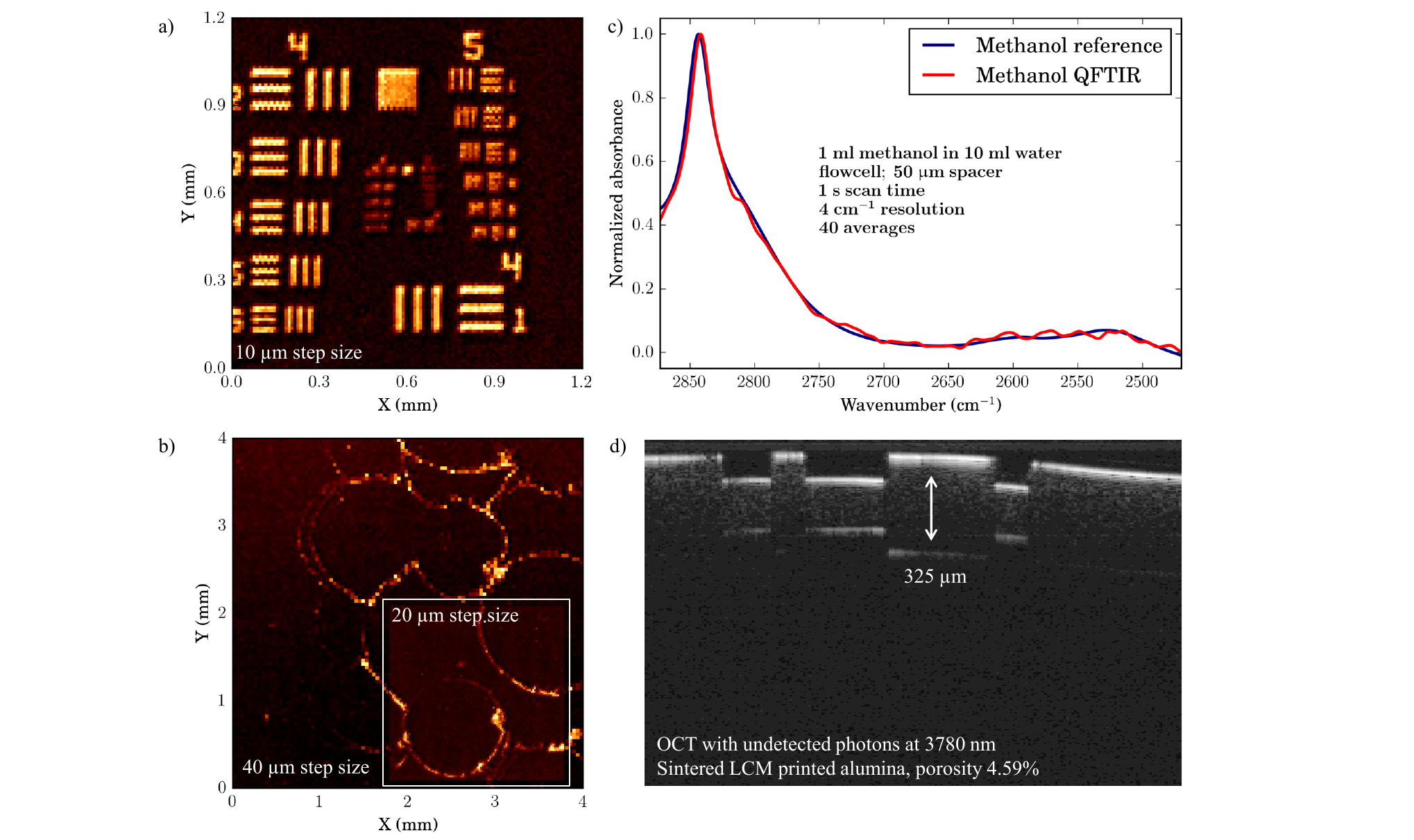}
\caption{Sensing with undetected photons implemented in various measurement modalities for applied non-destructive testing: IR microscopy (a,b), absorption transmission IR spectroscopy (c), and mid-IR OCT (d). (a) integrated reflected intensity of a 1951 USAF test target; (b) integrated absorbance of a pharmaceutical ingredient on a steel plate (45 \textmu g/cm$^2$); (c) normalized absorbance of methanol in water; (d) OCT with undetected photons B-scan of sintered LCM (lithography-based ceramic manufacturing) printed alumina with high porosity of 4.59~\%} \label{fig:mapping}
\end{figure*}

If the nonliear interferometer is used for QFTIR spectrometry or microscopy, time-domain interferograms are recorded by scanning the pump-signal arm, i.e., the SM mirror. An interferometer based on a HeNe laser tracks the movement of the SM that is used to linearize the interferograms. A custom Python software is used to process the recorded interferograms (linearization, apodization, zero filling, Fourier transform)~\cite{werefkin_ftir_2020} and to display the spectra on a graphical user interface. The software also controls the stage with the SM and two linear stages that move the sample for mapping applications (for both OCT and IR spectroscopic microscopy). For spectral-domain OCT measurements, the SM is fixed, and a different custom Python software is used~\cite{zorin2022oct}.

Mapping-based IR microscopy with undetected photons has significant potential to generate infrared microscopic images without requiring mid-IR illumination sources or IR detectors. 
As a first sample for the QFTIR microscopic mapping modality, a 1951 USAF test target was used to assess spatial resolution. The step size in x- and y-direction was 10~\textmu m. The integrated reflectance is shown in Fig.~\ref{fig:mapping}~(a). 

The linewidth of group 4, element 6 is 17.54~\textmu m and the linewidth of group 5, element 1 is 15.63~\textmu m which could still be resolved with the employed AC lens. Thus, the imaging modality demonstrates a high spatial resolution that is difficult to achieve with state-of-the-art classical FTIR systems, which possess a trade-off between spatial and spectral performance~\cite{Kilgus:18}. The pixelation artifact visible in element 2 of group 4 can be attributed to hysteresis in the stages (zig-zag scanning was used). 

\begin{table*}[t]
\centering
\caption{Comparison of QFTIR to a classical FTIR system.}
\label{table:benchmark}

\renewcommand{\arraystretch}{1.2}
\setlength{\tabcolsep}{8pt}

\begin{tabularx}{\textwidth}{
    @{}
    >{\raggedright\arraybackslash}X
    >{\centering\arraybackslash}X
    >{\centering\arraybackslash}X
    @{}
}
\toprule
\textbf{Performance metric}
&
\textbf{QFTIR (single-mode)}
&
\textbf{Classical FTIR (multi-mode)}
\\
\midrule

Sensing power
&
$60 \times 10^{-12}$\,W
&
$1.6 \times 10^{-3}$\,W
\\

SNR
&
36
&
962
\\

Shot-noise-limited SNR
&
199 {\footnotesize(5.5-fold)}
&
$2.1 \times 10^{5}$ {\footnotesize(217-fold)}
\\

SNR power scaling
&
$1.5 \times 10^{5}$\,mW$^{-1/2}$
&
764\,mW$^{-1/2}$
\\

\bottomrule
\end{tabularx}

\vspace{0.5em}

\begin{minipage}{0.94\textwidth}
\footnotesize
\raggedright
The SNR power scaling of the characterized QFTIR system is two to three orders of magnitude larger than that of the classical FTIR system. The results indicate that the QFTIR operates in the shot-noise-dominated regime. Its absolute SNR can therefore be increased by increasing the SPDC photon-pair generation rate.
\end{minipage}
\end{table*}

The next image shown in Fig.~\ref{fig:mapping}~(b) represents a real-life use case. A dried active pharmaceutical ingredient applied to a steel plate (with a concentration as low as 45~\textmu g/cm$^2$) was imaged. The integrated absorbance associated with the CH group (2700 cm$^{-1}$ - 3000 cm$^{-1}$) is shown in Fig.~\ref{fig:mapping}~(b). Typical coffee-stain rings of residue dried pharmaceutical ingredient can be observed, which are a common drying phenomenon in liquid solutions applied to flat surfaces. The two microscopic mapping images were scanned without averaging and, therefore, no hyperspectral analysis was performed. The measurement time per pixel was approximately 1 seconds.

The capabilities for IR absorbance measurements can be highlighted in routine transmission measurements: methanol in aqueous solution (9.1~vol$\%$) was measured in transmission through a liquid flow-cell with a 50~\textmu m spacer. For this measurement, the idler arm (in absence of the AC doublet) was terminated with a gold mirror. The liquid flow-cell was placed in the collimated beam path and 40 spectra were averaged. A spectral resolution of 4 cm$^{-1}$ and 1~s scan time were used as acquisition parameters. A reference measurement was performed with a state-of-the-art FTIR spectrometer (Bruker Invenio) with similar settings (but only 16 averages). The resulting absorbance spectra are shown in Fig.~\ref{fig:mapping}~(c). 

Spectral-domain OCT measurements were performed by coupling the signal photons to a spectrometer. As a sample, a sintered LCM alumina part with substantially high porosity (4.59~$\%$) was used (Fig.~\ref{fig:mapping}~(d)). It should be noted that with classical commercially available OCT platforms, the interface at 325~\textmu m below the sample surface could not be detected due to high scattering losses.

In order to assess the applied capabilities of the QFTIR modality, several measures have been benchmarked against a state-of-the-art FTIR spectrometer~\cite{Gattinger2025}. These values are collected in Table~\ref{table:benchmark}. The probing idler power of 60 pW (after the first pass) was evaluated by measuring the optical power in the signal domain (around 565~pW after double pass) with a suitable calibrated photonic sensor; the idler photon rate was calculated using energy conservation. This confirmed an SPDC conversion efficiency on the order of $10^{-9}$. The signal-to-noise ratio (SNR) was evaluated for the standard deviation of 100~\% lines (fluctuations between two blank spectra) that were zero-centered. The shot-noise limited SNR is the standard quantum limit; i.e. the highest achievable SNR. It was calculated by taking into account the photon shot noise for a given photon rate, the efficiency of the interferometer (visibility), as well as the quantum efficiency of the detector~\cite{Lindner:21}. In the given configuration, the theoretical shot-noise-limited SNR was estimated to be 199, whereas the measured SNR was 36. Thus the nonlinear interferometer was evaluated to be 5.5 times off the shot-noise-limited performance. This offset can be attributed to drifts and vibrations in the laboratory setup and errors in the measurement of the visibility and the quantum efficiency of the detector. However, if the same formalism is applied to the reference classical FTIR the resulting theoretical shot-noise-limited SNR of $2.1\times 10^5$ is 217 times off the measured SNR of 962. This illustrates the fundamental limitation of classical IR instruments by noise sources such as thermal and 1/$f$ noise. The last metric that was compared is the power dependence of the SNR. It can be calculated by multiplying the SNR by the square root of the inverse sensing power. For the interferometer under investigation this yields $1.5 \times 10^5$~mW$^{-1/2}$ whereas it is 764~mW$^{-1/2}$ for the classical FTIR. This indicates that the SNR per mW of power is two to three magnitudes higher for the QFTIR, thus highlighting that the technology could eventually be competitive to classical IR spectroscopy, given that the sources of correlated photons can be made brighter. It should also be noted that the classical FTIR in this comparison is a multi-mode system, whereas the QFTIR is single-mode based.

\section{Summary and conclusion}\label{sec_conc}

Metrology with undetected photons is an emerging field of quantum sensing with significant potential for numerous practical applications. We demonstrated a nonlinear Michelson-type interferometer and used it for QFTIR spectroscopy, microscopic mapping, and mid-IR spectral-domain OCT with undetected photons. The latter can already compete with state-of-the-art systems not only performance-wise but also in terms of technical difficulty and price. Classical mid-IR OCT is still an active research field and it requires costly and technically challenging mid-IR sources. It is interesting to point out that a mid-IR supercontinuum source, which is needed for such a system, is a factor of 2-3 more expensive than the total OCT with undetected photons setup presented in this work. Another technical limitation is given by mid-IR detectors, which are dominated by thermal and 1/f noise. Fourier-domain OCT requires fast detector arrays with a sufficient number of pixels to ensure high axial resolution and reasonable probing depth. Such detectors are scarce and expensive in the mid-IR domain. Alternatively, time-encoded approaches can be used, which also rely on expensive and noisy mid-IR supercontinuum sources~\cite{Zorin2021}. 
In contrast, the presented approach renders both direct mid-IR sources and detectors obsolete.

An interesting and noteworthy feature of metrology with undetected photons is the ultra-low power impinging on the samples, which is 60~pW for the demonstrated system. This could be potentially useful for the investigation of delicate samples from art to biology, as well as concealed detection schemes. This represents a potential paradigm shift in mid-IR sensing that enables shot-noise-limited performance at ultra-low optical powers below the thermal background.
Moreover, the linear dependence of the signal and idler photon rates on the pump power in the low-gain regime enables the use of significantly more cost-effective components. For instance, transitioning to a substantially cheaper pump laser with four times lower optical power results in only a twofold reduction in SNR, as the system operates in the shot-noise-dominated regime. Such sensing solutions are currently in development. 

Nevertheless, efforts have to be made to increase the brightness of SPDC sources to improve the SNR for QFTIR spectroscopy as well as OCT with undetected photons. Due to the quasi-shot-noise-limited performance of the detectors used in IR metrology with undetected photons, these techniques could ultimately outperform classical systems in terms of sensitivity. Some efforts have already been demonstrated~\cite{Lindner2023}, and other works go into the direction of increasing the spectral bandwidth of these systems and pushing it towards longer wavelengths in the IR domain~\cite{Tashima2024,Cheng2026}, which is essential for IR spectroscopy. 

\bibliography{main}

\end{document}